\documentclass[letter,linenoaa]{aa}
\usepackage{amssymb,amsmath}
\usepackage{graphicx}
\usepackage{xcolor,natbib}
\usepackage{multirow}
\usepackage[T1]{fontenc}
\usepackage{ae,aecompl}
\usepackage{newtxtext, newtxmath}
\usepackage{float}
\usepackage{subfigure}
\usepackage{longtable}
\usepackage{enumitem}
\usepackage{longtable,listings}
\usepackage[flushleft]{threeparttable}
\usepackage{parcolumns}
\usepackage{hyperref}

\def\dif{\mathop{}\!\mathrm{d}}

\begin{document}

\title{Quasi-periodic oscillations in optical color evolutions to support sub-pc binary black hole systems in broad 
line active galactic nuclei}

\titlerunning{QPOs in optical color evolutions}

\author{XueGuang Zhang}

\institute{Guangxi Key Laboratory for Relativistic Astrophysics, School of Physical Science and Technology, GuangXi 
University, No. 100, Daxue Road, Nanning, 530004, P. R. China \ \ \ \email{xgzhang@gxu.edu.cn}}

\abstract
{Optical quasi-periodic oscillations (QPOs) with periodicity around hundreds to thousands of days have been accepted 
as an efficient indicator for sub-pc binary black hole systems (BBHs) in broad line active galactic nuclei (BLAGN). 
However, considering intrinsic variability (red noises) of BLAGN, it is still an open question on physical origin of 
detected optical QPOs from AGN variability or truly from sub-pc BBHs. Here, a simple method is proposed to support 
optical QPOs related to sub-pc BBHs by detecting QPOs in time dependent optical color evolutions of BLAGN. Periodic 
variations of obscurations are expected on optical light curves related to sub-pc BBHs, but there should be non-periodic 
variable obscurations on optical light curves in normal BLAGN. Through simulated optical light curves for intrinsic 
AGN variability by Continuous AutoRegressive process with time durations around 2800days (similar as time durations 
of light curves in ZTF), the probability is definitely smaller than $3.3\times10^{-7}$ that QPOs can be detected in 
the corresponding optical color evolutions, but about $3.1\times10^{-2}$ that optical QPOs with periodicity smaller 
than 1400days (at least two cycles) can be detected in the simulated single-band light curves. Therefore, confidence 
level is definitely higher than 5$\sigma$ to support the QPOs in color evolutions not related to intrinsic AGN 
variability but truly related to sub-pc BBHs. In the near future, the proposed method can be applied for searching 
reliable optical QPOs in BLAGN through multi-band light curves from the ZTF and the upcoming LSST.
}


\maketitle

\section{Introduction}

	Sub-pc binary black hole systems (BBHs)	have been commonly expected as natural products of galaxy mergers 
and evolution \citep{bb80, sr98, md06, se14, mj22, ag24, cm25}. Considering periodic orbital rotations of two BH 
accreting systems in sub-pc BBHs, long-standing optical quasi-periodic oscillations (QPOs) related to periodic 
obscurations have been widely accepted as the efficient indicator for sub-pc BBHs. Until now, through optical QPOs 
with periodicity around hundreds to thousands of days, there are around 300 sub-pc BBHs reported in broad line 
active galactic nuclei (BLAGN), such as the samples of candidates in \citet{gd15, cb16, fw25} and the individual 
candidates in \citet{gd15a, lg15, kp19, zb16, ss20, ky20, lw21, zh22a, zh22b, zh23a, zh23b, ap24, lz25, zh25a}, etc.

	Although more and more long-standing optical QPOs have been reported in BLAGN, whether the detected QPOs 
were intrinsically related to sub-pc BBHs is also an open question. More specifically, red noises, traced by 
intrinsic AGN variability as one of fundamental characteristics of AGN \citep{ww95, uw97, bg18, bp25}, can lead 
to mathematically determined fake optical QPOs. Effects of red noises on optical QPOs have been firstly shown in 
\citet{vu16}. Then, in our recent works in \citet{zh23a, zh23b, lz25, zh25a}, such effects of red noises have been 
discussed, especially on corresponding confidence levels to support the detected optical QPOs not related to 
intrinsic AGN variability through mathematical simulations.

	Due to apparent and inevitable effects of red noises on optical QPOs in BLAGN, it is a challenge 
to propose a method applied to improve the confidence levels for detected optical QPOs in BLAGN, which is our main 
objective. Here, not through a single-band optical light curve but through optical color evolutions between two-bands 
optical light curves, an improved method is proposed to support sub-pc BBHs with confidence levels higher than 
5$\sigma$ in this manuscript through simulated results.

	For a single-band optical light curve in BLAGN, as discussed in \citet{kbs09, koz10, mac10, mw24}, it can 
be mathematically described by the widely accepted damped random walk (DRW) process or the first order continuous 
autoregressive (CAR) process with two basic process parameters of the intrinsic variability timescale $\tau$ and 
intrinsic variability amplitude $\sigma_*$. Certainly, improved higher-order continuous-time autoregressive moving 
average (CARMA) models have been proposed in \citet{kbs14, yr22, kg24} to describe AGN variability with more subtle 
features. However, due to more free model parameters that are difficult to constrain, the CARMA models are not 
convenient models to do our following mathematical simulations.

	Based on single-band optical light curve created by DRW/CAR process in BLAGN, the other one optical band 
light curve can be simply simulated, after considering time lags and probably different emission structures of 
two-bands optical emissions. Time lags around several days between two-bands optical light curves have been 
discussed for more than three decades as discussed in \citet{ea96, sd05, eg15, ym20, kp21, nh22, wl25}. Furthermore, 
considering the same source of ionization variability, similar variability patterns are accepted for optical light 
curves in different optical bands but convolved with a kernel function for considerations of probably different 
structure information, similar as done in \citet{zk16, mw24}.

	In this manuscript, considering no variable obscurations on intrinsic AGN variability but apparent time 
lags between two-bands optical light curves, an improved method through detecting QPOs in optical color (difference 
between two-bands optical light curves) evolutions is mainly discussed to support sub-pc BBHs with higher confidence 
levels. The manuscript is organized as follows. Section 2 presents the main hypotheses and results. The main 
conclusions are given in Section 3. Throughout the manuscript, we have adopted the cosmological parameters of 
$H_{0}$=70 km s$^{-1}$ Mpc$^{-1}$, $\Omega_{m}$=0.3, and $\Omega_{\Lambda}$=0.7.

\begin{figure}
\centering\includegraphics[width = 8cm,height=5cm]{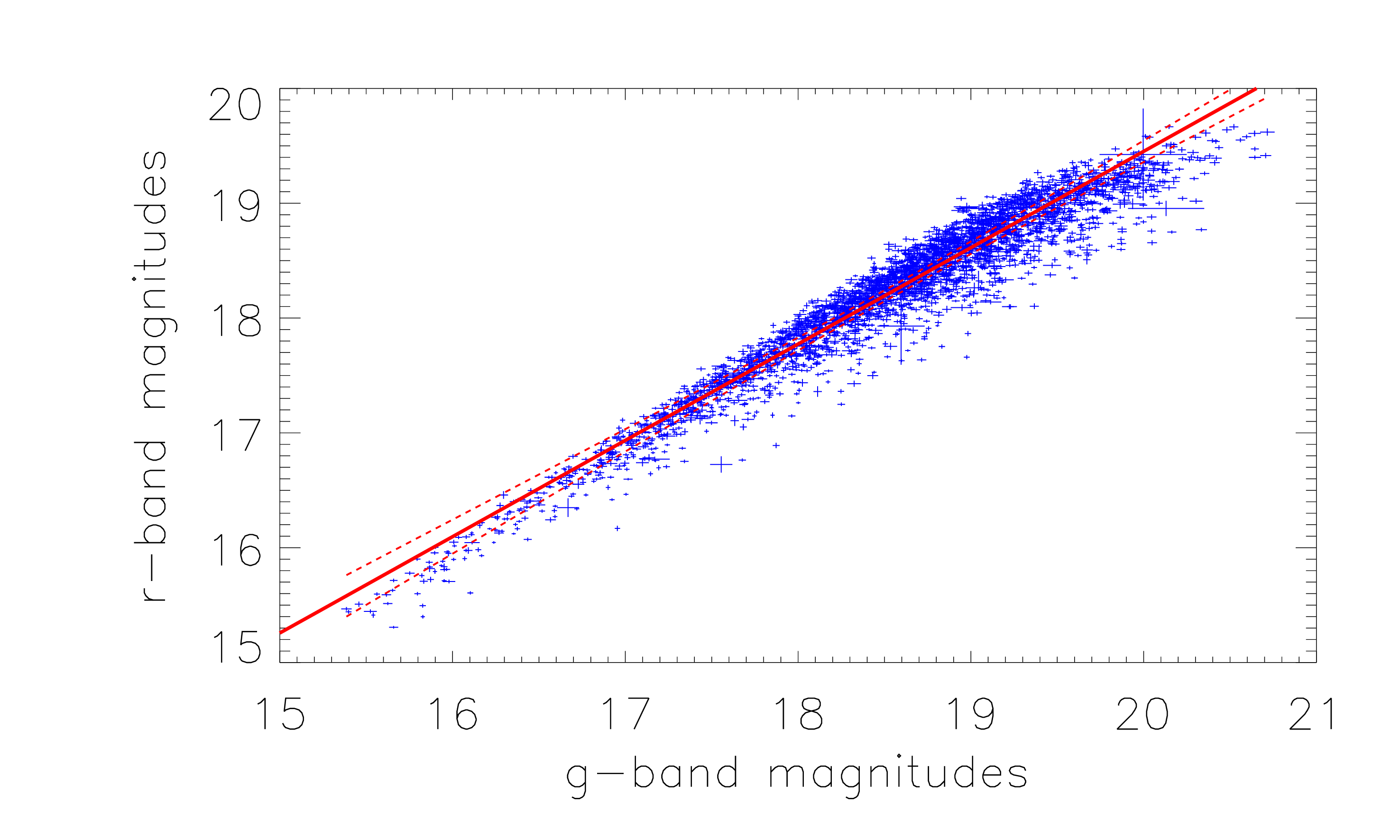}
\caption{On the correlation between SDSS g- and r-band apparent magnitudes of 3530 low redshift quasars. Solid 
and dashed red lines show the best fitting results and corresponding 5$\sigma$ confidence bands.}
\label{mag2}
\end{figure}

\begin{figure*}
\centering\includegraphics[width = 18cm,height=10cm]{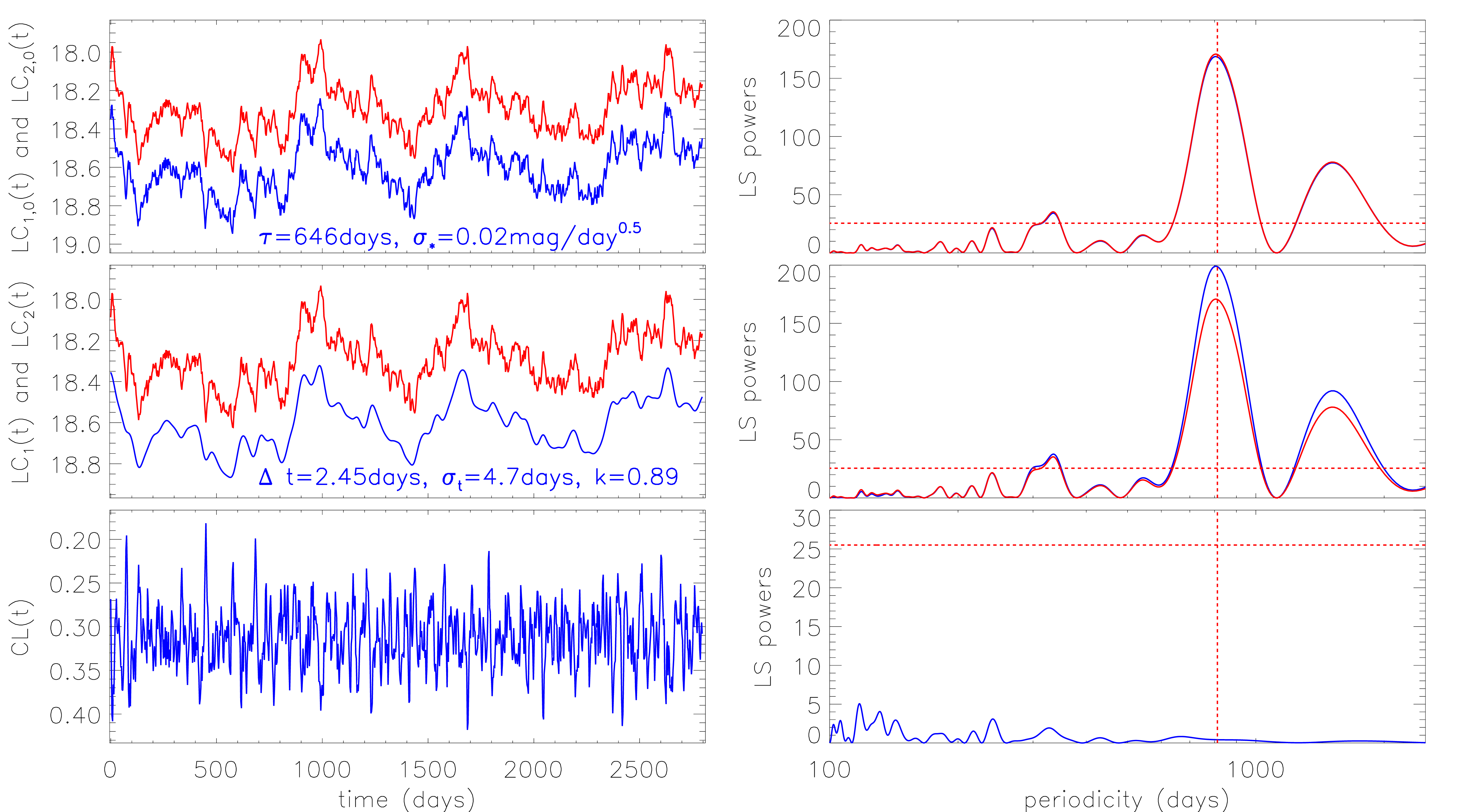}
\caption{Left panels show the light curves of $LC_{1,0}(t)$ (in blue) and $LC_{2,0}(t)$ (in red) (top left panel),
$LC_{1}(t)$ (in blue) and $LC_{2}(t)$ (in red) (middle left panel) and $Color(t)$ (bottom left panel). Right panels
show the corresponding LS powers. In top right panel (middle right panel), solid line in blue and in red show the
results through $LC_{1,0}(t)$ ($LC_{1}(t)$) and $LC_{2,0}(t)$ ($LC_{2}(t)$), respectively. In each right panel,
horizontal dashed red line marks the 5$\sigma$ significance level (determined by the input false alarm probability
to be $3\times10^{-7}$), vertical dashed red line marks the position of periodicity around 812days determined in
$LC_{1}(t)$. }
\label{lc2}
\end{figure*}

\section{Hypotheses, Results and Discussions}

\subsection{Main results through the simulated light curves}

	As described in Introduction, variability of optical color evolutions from two-bands light curves are 
mainly considered. Therefore, among the current public sky survey projects, the Zwicky Transient Facility (ZTF) 
\citep{ref8, ref9} is the most suitable choice at current stage, which provides optical g/r-bands light curves 
with time durations around 2800days.  

	Before to create artificial optical two-bands light curves of BLAGN, the mean values of two-bands apparent 
magnitudes can be simply determined by the correlation between Sloan Digital Sky Survey (SDSS) g-band and r-band 
magnitudes ($m_g$, $m_r$) of low redshift ($z<0.35$) 3530 quasars listed in \citet{sh11}. As shown in Fig.~\ref{mag2}, 
there is a strong linear correlation with Spearman Rank correlation coefficient 0.96 ($P_{null}<10^{-10}$). 
Considering uncertainties in both coordinates, through the Least Trimmed Squares regression technique \citep{cap13}, 
the correlation can be described as 
\begin{equation}
m_r~=~(2.672\pm0.057)~+~(0.839\pm0.003)\times~m_g,
\end{equation}
with RMS scatter about 0.154. Although the apparent magnitudes are collected from SDSS not from ZTF, there are few 
effects on our following results. Then, the equation can be applied to determine the apparent magnitudes of the 
following artificial optical two-bands light curves by the following two steps.

	First, through the CAR process \citep{kbs09}, a single-band optical light curve $LC_{1,0}(t)$ can be 
created as
\begin{equation}
LC_{1,0}(t)~=~\frac{-1}{\tau}LC_{1,0}{t}\dif t+\sigma_{*}\sqrt{\dif t}\epsilon(t)~+~m_1,
\end{equation}
with $\epsilon(t)$ as a white noise process with zero mean and variance equal to 1. The corresponding time duration 
of $t$ is about 2800days and the time step is about 3days, similar as the time durations and time steps of high 
quality light curves in ZTF. Here, the time $t$ is create by 993 random values from 0 to 2800 (corresponding IDL code, 
t=randomu(seed, 993) * 2800) \& t=t(sort(t)) \& t=t(rem\_dup(t)). The CAR process parameters are randomly collected 
from 100days to 1500days for $\tau$ and from 0.003${\rm mag/day^{0.5}}$ to 0.04${\rm mag/day^{0.5}}$ for $\sigma_*$, 
which are common value for BLAGN as shown in \citet{kbs09, koz10, mac10}. Then, the mean apparent magnitudes $m_1$ 
(assumed in g-band) are randomly collected from 16mags to 20mags, as shown in Fig.~\ref{mag2}.

	Second, light curve $LC_2(t)$ from the other optical band can be created as follows. Considering the results 
in Fig.~\ref{mag2}, the mean apparent magnitude $m_{2}$ (assumed in r-band) is collected by 
\begin{equation}
	m_2~=~(2.672~+~k_1)~+~(0.839~+~k_2)\times~m_1,
\end{equation},
with $k_1$ and $k_2$ as random values determined by normal distributions centered at zero and with second moments 
to be the corresponding uncertainties of 0.057 and 0.003 determined in Equation (1). Considering 
time lag $\Delta t$ between g-band and r-band with random values from 0days to 5days (the value large enough) as 
more recent discussions in \citet{nh22, wl25}, the delayed light curve can be described as 
\begin{equation}
	LC_{2,0}(t)~=~LC_{1,0}(t~-~\Delta t)~-~m_1~+~m_2,
\end{equation}.
Then, considering probably different structure information of emission regions for the two-bands optical light 
curves, a Gaussian convolution function $G(\sigma_t)$ with second moment $\sigma_t$ randomly from 0days to 5days 
is applied as
\begin{equation}
\begin{split}
	&k~\in~[0,~1] \\
	&LC_1(t)~=~LC_{1,0}(t)\ \ \ \ \ \ (k<0.5) \\
        &LC_2(t)~=~LC_{2,0}(t)~\otimes~G(\sigma_t)   \ \ \ \ \ \ (k<0.5) \\
        &LC_2(t)~=~LC_{2,0}(t)\ \ \ \ \ \ (k\ge0.5) \\
        &LC_1(t)~=~LC_{1,0}(t)~\otimes~G(\sigma_t)   \ \ \ \ \ \ (k\ge0.5)
\end{split},
\end{equation}
with $k$ as a random value from 0 to 1. Then, the color evolution
\begin{equation}
CL(t)~=~LC_1(t)~-~LC_2(t)
\end{equation}
can be calculated.

	Based on simulated $LC_1(t)$ and $CL(t)$, it is interesting to check whether probably fake QPOs can be 
detected in the artificial light curves tightly related to AGN variability (red noises). To repeat the two steps 
above three million times, QPOs can be checked through the $LC_1(t)$ and $CL(t)$, which will provide clues to 
determine probability for detecting fake QPOs in the corresponding light curves. Here, the commonly accepted 
Lomb-Scargle (LS) periodogram \citep{vj18} has been accepted to detect QPOs. Fig.~\ref{lc2} shows examples of 
$LC_{1,0}(t)$ and $LC_{2,0}(t)$, $LC_1(t)$ and $LC_2(t)$ and $CL(t)$ and corresponding LS powers. As the shown 
examples, it is clear that the CAR created artificial light curves can lead to detected fake QPOs with 
significance level higher than 5$\sigma$ in single-band light curves. However, such fake QPOs cannot be detected 
in the corresponding $CL(t)$, providing clues that probability of detecting fake QPOs in $CL(t)$ should be 
significantly smaller than in single-band optical light curve. 

	Among the three million ($N_t=3d6$) simulated cases, two probabilities of detecting fake QPOs are 
determined. The first probability $P_{1}$ is for detecting QPOs in $LC_1(t)$ with significance level higher than 
5$\sigma$ and the corresponding periodicity $T_{P1}$ smaller than 1400days (to support the artificial $LC_1(t)$ 
including at least two cycles), leading to $N_1=94527$ $LC_1(t)$ to be collected. Therefore, through 
intrinsic AGN variability, fake optical QPOs can be detected, and the corresponding probability $P_1$ is about 
$N_1/N_t~\sim~3.1\times10^{-2}$. Meanwhile, the probability $P_{12}\sim3.1\%$ can also be determined 
for detecting QPOs in $LC_2(t)$ after considering the same criteria applied for $P_1$, due to $N_2=94579$ 
$LC_2(t)$ to be collected. There are the same probabilities for detecting fake QPOs in the simulated $LC_1(t)$ and 
$LC_2(t)$. Furthermore, for the 94527 $LC_1(t)$ including fake QPOs, Fig.~\ref{pars} in the Appendix 
shows distributions of the parameters of $\tau$, $\sigma$, $\Delta t$ and $\sigma_t$ applied in the procedure 
above, indicating that smaller $\tau$ should favor the detecting fake QPOs in the simulated light curves.
Moreover, in \citet{zh25a} for optical QPOs in PG 1411, the probability is $P_0\sim4.8\times10^{-4}$, very smaller 
than $P_1$, for detecting fake QPOs through the CAR process simulated light curves. However, in \citet{zh25a}, 
additional criteria have been applied that periodicity from 450days to 650days and sine function described 
simulated light curves. If accepted the two additional criteria, the number should be $N_1\sim2193$, leading 
to the probability $P_{10}\sim7.3\times10^{-4}$ similar as $P_0$. The small difference between $P_{10}$ and 
$P_0$ is probably due to different qualities of simulated light curves, different time steps and different time 
durations.

	The second probability $P_{2}$ is for detecting QPOs in $CL(t)$ with significance level higher than 
5$\sigma$ and also with the corresponding periodicity $T_{P2}$ similar to the $T_{P1}$ detected in $LC_1(t)$, 
leading to no one $CL(t)$ to be collected. Here, we simply accepted the similar periodicities in $LC_1(t)$ 
and in $CL(t)$, if $\frac{T_{P1}}{T_{P2}}$ between 0.85 and 1.15. Therefore, through the AGN variability 
expected color evolution $CL(t)$, no fake QPOs can be detected, and the corresponding probability is about 
$P_{2}~<~1/N_t~\sim~3.3\times10^{-7}$, about $9.5\times10^4$ times smaller than $P_1$. Based on the determined 
$P_2$, we can safely state that once QPOs detected in $CL(t)$, the confidence level should be higher than 
5$\sigma$ to support that the detected QPOs are not related to intrinsic AGN variability. Therefore, to detect 
QPOs in $CL(t)$ should provide stable evidence to support QPOs not related to intrinsic AGN variability but 
truly related to sub-pc BBHs.

\begin{figure}
\centering\includegraphics[width = 8cm,height=12cm]{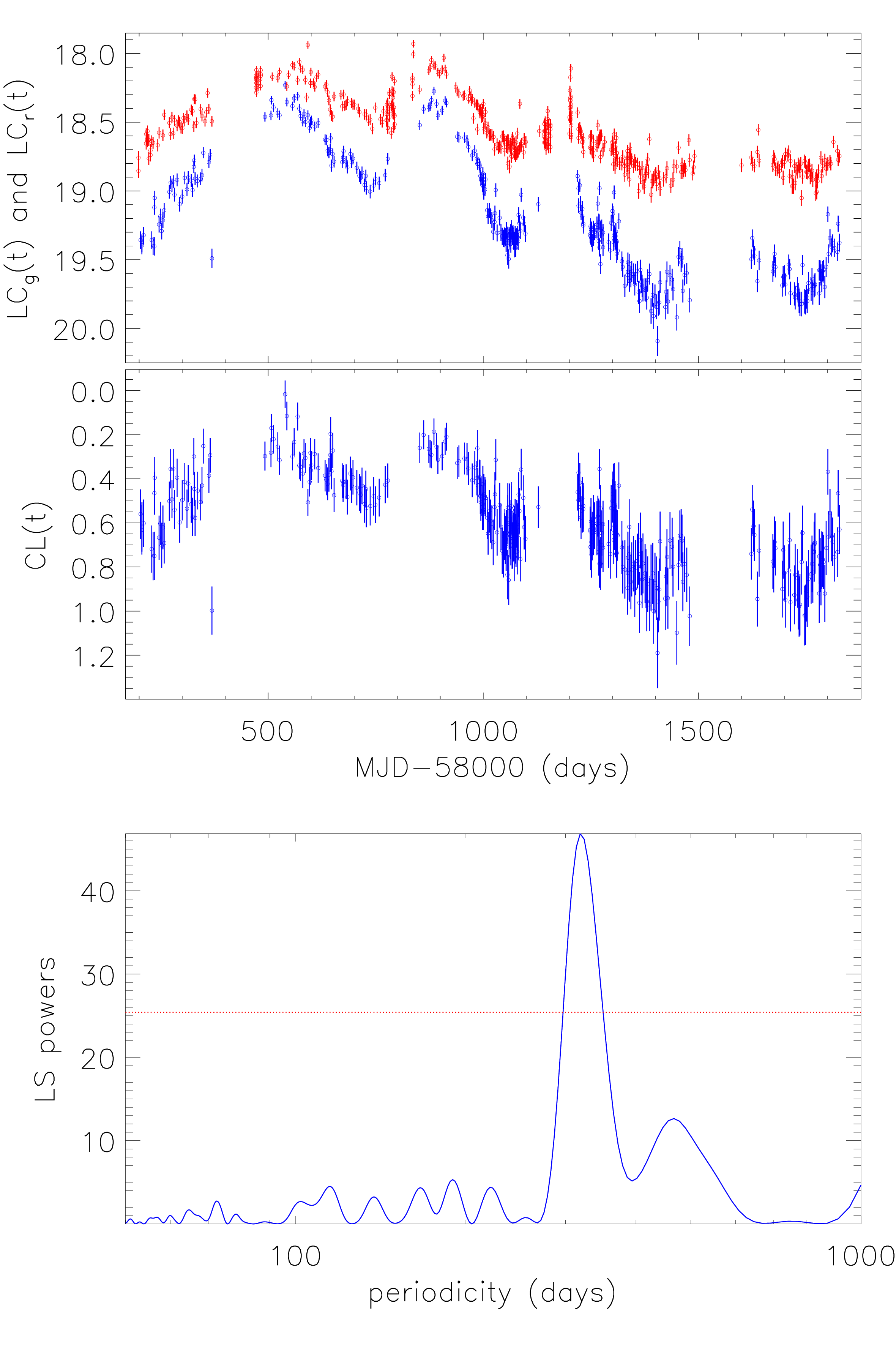}
\caption{Top panel shows the ZTF g-band (blue symbols) and r-band (red symbols) light curves of SDSS J1609+1756. 
Middle panel shows the $CL(t)$. Bottom panel shows the determined LS powers of $CL(t)$, with horizontal dashed 
red line as 5$\sigma$ significance level.}
\label{J1609}
\end{figure}

\subsection{Application in the SDSS J1609+1756}
	The discussed results above are based on the given time information of light curves and randomly collected 
process parameters. It is necessary to test such results in long-term optical two-bands light curves of a real 
BLAGN. Then, we checked the reported optical QPOs with periodicity about 340days in the blue quasar SDSS J1609+1756 
in \citet{zh23a} through long-term ZTF light curves. The g/r-band light curves of $LC_g(t)$ and $LC_r(t)$ and the 
$CL(t) = LC_g(t) - LC_r(t)$ are shown in top two panels of Fig.~\ref{J1609}. The corresponding LS powers of $CL(t)$ 
are shown in the bottom panel of Fig.~\ref{J1609}, leading to periodicity about 320days in $CL(t)$. Then, based on 
the time information $t$ of g-band light curve and the determined $\tau\sim140days$ and 
$\sigma_*\sim0.035{\rm mag/day^{0.5}}$ reported in \citet{zh23a}, three million simulations have been done as what 
have recently done above, still leading to no one case collected with apparent QPOs in $CL(t)$. Therefore, relative 
to the probability about 0.26\% in \citet{zh23a} for detecting fake QPOs in CAR procedure simulated single-band 
optical light curves, the probability is apparently smaller than $3.3\times10^{-7}$ ($~<~1/3d6$) to detect QPOs 
in $CL(t)$. In other words, the confidence level is higher than 5$\sigma$ for the optical QPOs in the blue quasar 
SDSS J1609+1756.

\section{Conclusions}

	Considering periodic variations in obscurations on light curves related to sub-pc BBHs but no variable 
obscurations on light curves of normal BLAGN, we propose a method to support detected QPOs not related to red 
noises (AGN intrinsic variability) but truly related to sub-pc BBHs, through reliable periodic features in time 
dependent optical color evolutions. Among $3\times10^6$ simulated cases of $CL(t)$ by artificial two-bands optical 
light curves for intrinsic AGN variability by the CAR process, the probability for detecting QPOs in $CL(t)$ is 
definitely smaller than $3.3\times10^{-7}$, which is about $9.5\times10^4$ times smaller than the probability 
for detecting QPOs in single-band light curves. Therefore, confidence level higher than 5$\sigma$ can be safely 
accepted for optical QPOs in a BLAGN, if there is QPOs in corresponding optical color evolutions. The method has 
been applied in the blue quasar SDSS J1609+1756, to support the reported optical QPOs related to sub-pc BBHs. 
The method proposed in this manuscript can be efficiently applied for searching for reliable optical QPOs in 
BLAGN through multi-band light curves from the ZTF and the upcoming project of Legacy Survey of Space and Time 
(LSST) in the near future.

\begin{acknowledgements}
Zhang gratefully acknowledge the anonymous referee for giving us constructive comments and suggestions 
to greatly improve the paper. Zhang gratefully thanks the kind financial support from GuangXi University and the 
kind grant support from NSFC-12173020, 12373014 and from the Guangxi Talent Programme (Highland of Innovation Talents). 
This article is kindly dedicated to celebrating the 60th birthday of my most respected supervisor, Prof. Wang 
TingGui. This manuscript has made use of the data from ZTF (\url{https://www.ztf.caltech.edu}), and use of the IDL 
code of lts\_linefit.pro (\url{https://users.physics.ox.ac.uk/~cappellari/}) and scargle.pro 
(\url{http://astro.uni-tuebingen.de/software/idl/aitlib/timing/scargle.html}).
\end{acknowledgements}

\appendix
\section{Parameter distributions of the $LC(t)$ including fake QPOs}

\begin{figure}
\centering\includegraphics[width = 8cm,height=17cm]{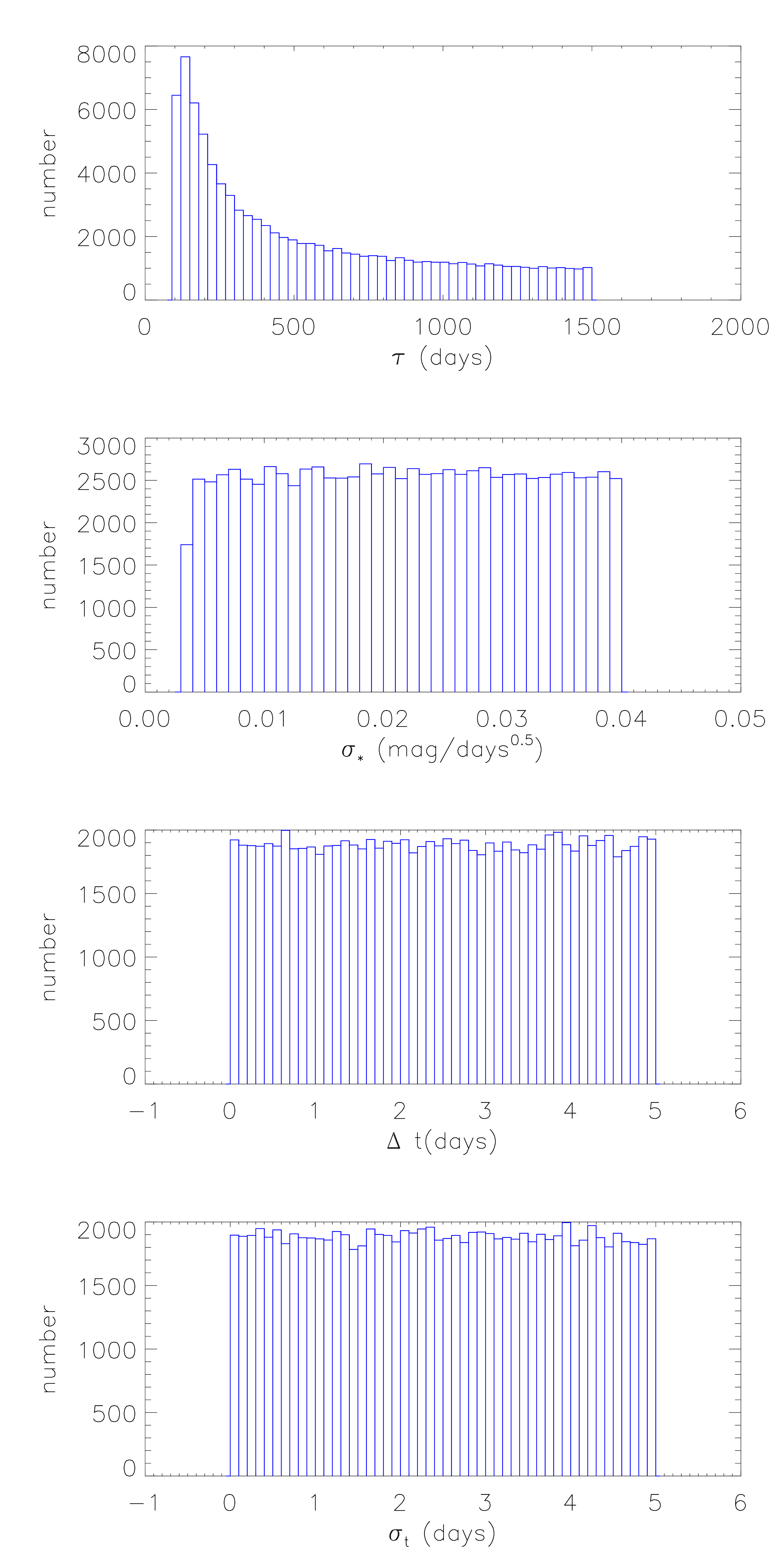}
\caption{Distributions of the $\tau$, $\sigma$, $\Delta t$ and $\sigma_t$ of the $N_1=94527$ artificial 
	$LC_1(t)$ including fake QPOs.}
\label{pars}
\end{figure}

	For the $N_1=94527$ artificial $LC_1(t)$ including fake QPOs, the distributions of the $\tau$, $\sigma$, 
$\Delta t$ and $\sigma_t$ are shown in the Fig.~\ref{pars}. It is clear that smaller $\tau$ should favor the 
detecting fake QPOs in simulated light curves. And the other three parameters have few effects on detecting fake 
QPOs. Meanwhile, for the $N_2=94579$ $LC_2(t)$ including fake QPOs, totally similar results can be confirmed. 
And there are no re-plotted results for the parameters of the $N_2=94579$ $LC_2(t)$ in Fig.~\ref{pars} due to 
totally overlapped results in the plots.

\end{document}